\documentclass[apjl]{emulateapj}
\pdfoutput=1

\usepackage[fleqn]{amsmath} 
\usepackage{amsfonts} 
\usepackage{amstext}
\usepackage{amssymb} 
\usepackage[colorlinks=true, citecolor=blue, linkcolor=blue, breaklinks=true]{hyperref}
\usepackage{natbib}
\newcommand{\hi } {{\rm H}\,{\small\rm I} \,}
\newcommand{\hiA} {{\rm H}\,{\small\rm I}}

\newcommand{\sdyn}{$\Sigma_{\rm dyn}(0)$}
\newcommand{\sstar}{$\Sigma_{\star}(0)$}
\newcommand{\ml}{$\Upsilon_{[3.6]}$}

\begin{document}
\nocite{*}

\title{The relation between stellar and dynamical surface densities in the central regions of disk galaxies}
\author{Federico Lelli$^{1}$, Stacy S. McGaugh$^{1}$, James M. Schombert$^{2}$, and Marcel S. Pawlowski$^{1}$}
\affil{$^{1}$Department of Astronomy, Case Western Reserve University, Cleveland, OH 44106, USA; federico.lelli@case.edu\\
$^{2}$Department of Physics, University of Oregon, Eugene, OR 97403, USA}

\begin{abstract}
We use the SPARC (Spitzer Photometry \& Accurate Rotation Curves) database to study the relation between the central surface density of stars \sstar\ and dynamical mass \sdyn\ in 135 disk galaxies (S0 to dIrr). We find that \sdyn\ correlates tightly with \sstar\ over 4 dex. This central density relation can be described by a double power law. High surface brightness galaxies are consistent with a 1:1 relation, suggesting that they are self-gravitating and baryon dominated in the inner parts. Low surface brightness galaxies systematically deviate from the 1:1 line, indicating that the dark matter contribution progressively increases but remains tightly coupled to the stellar one. The observed scatter is small ($\sim$0.2 dex) and largely driven by observational uncertainties. The residuals show no correlations with other galaxy properties like stellar mass, size, or gas fraction.
\end{abstract}


\section{Introduction}\label{sec:intro}

Several lines of evidence suggest that the stellar and total surface densities of disk galaxies are closely linked. The inner rotation curves of low surface brightness (LSB) galaxies rise more slowly than those of high surface brightness (HSB) ones \citep{deBlok1996a, Verheijen1997}, indicating that low stellar densities correspond to low dynamical densities \citep{deBlok1996b}. \citet{Lelli2013} find that the inner slope of the rotation curve $S_0$ (extrapolated for $R\rightarrow0$) correlates with the central surface brightness $\mu_0$ over 4 dex. This scaling relation has been independently confirmed by \citet{ErrozFerrer2016}, who show that other structural parameters (stellar mass, bulge-to-disk ratio, bar strenght) do \textit{not} correlate with $S_0$ as tightly as $\mu_{0}$. The inner slope $S_0$ scales as the square root of the central dynamical surface density \sdyn, hence the $S_0-\mu_0$ relation provides key information on the ratio between baryons and dark matter (DM) in galaxy centers \citep{Lelli2014}.

In this Letter we explore another method to estimate \sdyn. \citet{Toomre1963} provides the relation between the central surface density and the rotation curve  $V(R)$ of self-gravitating disks. His Equation 16 states
\begin{equation}\label{Eq:Toomre}
 \Sigma_{\rm dyn}(0) = \dfrac{1}{2\pi G} \int_{0}^{\infty} \dfrac{V^{2}(R)}{R^{2}}dR,
\end{equation}
where $G$ is Newton's constant. This formula holds as long as the baryonic disk is nearly maximal (Sect.\,\ref{sec:SigmaDyn}). \sdyn\ has key advantages over $S_0$: (i) it is independent of fitting procedures or extrapolations for $R \rightarrow 0$, and (ii) the disk thickness can be easily taken into account.

We consider 135 galaxies from the SPARC (Spitzer Photometry and Accurate Rotation Curve) database \citep[][hereafter Paper I]{Lelli2016a}. These objects have both high-quality \hi rotation curves and $Spitzer$ [3.6] surface photometry. The availability of [3.6] images is a major improvement over previous studies \citep{Lelli2013} since the near-IR surface brightness provides the best proxy to the central stellar surface density \sstar. We find that \sdyn\ tightly correlates with \sstar\ over 4 dex (Figure\,\ref{fig:Central}), even for LSB galaxies that appear not to be self-gravitating.

\section{Data Analysis}

We use galaxies from the SPARC database (Paper I). SPARC spans the widest possible range for disk galaxies: morphologies from S0 to dIrr, luminosities from $\sim$10$^7$ to $\sim$10$^{12}$ $L_{\odot}$, effective surface brightnesses from $\sim$5 to $\sim$5000 $L_{\odot}$~pc$^{-2}$, effective radii from $\sim$0.3 to $\sim$15 kpc, rotation velocities from $\sim$20 to $\sim$300 km~s$^{-2}$, and gas fractions from $\sim$0.01 to 0.95. In Paper\,I we describe the analysis of [3.6] images and the rotation curve data. We also define a quality flag: $Q=1$ indicates galaxies with high-quality \hi data or hybrid \hiA/H$\alpha$ rotation curves (99 objects), $Q=2$ indicates galaxies with minor asymmetries or \hi data of lower quality (64 objects), and $Q=3$ indicates galaxies with major asymmetries, strong non-circular motions, or off-sets between stellar and \hi distributions (12 objects). We exclude objects with $Q=3$ since the rotation curves do not necessarily trace the equilibrium gravitational potential. We also exclude face-on ($i < 30^{\circ}$) and edge-on ($i > 85^{\circ}$) galaxies due to uncertain corrections on the rotation velocities and central surface brightnesses, respectively. Our final sample consists of 135 galaxies.

\begin{figure}[thb]
\centering
\includegraphics[width=0.475\textwidth]{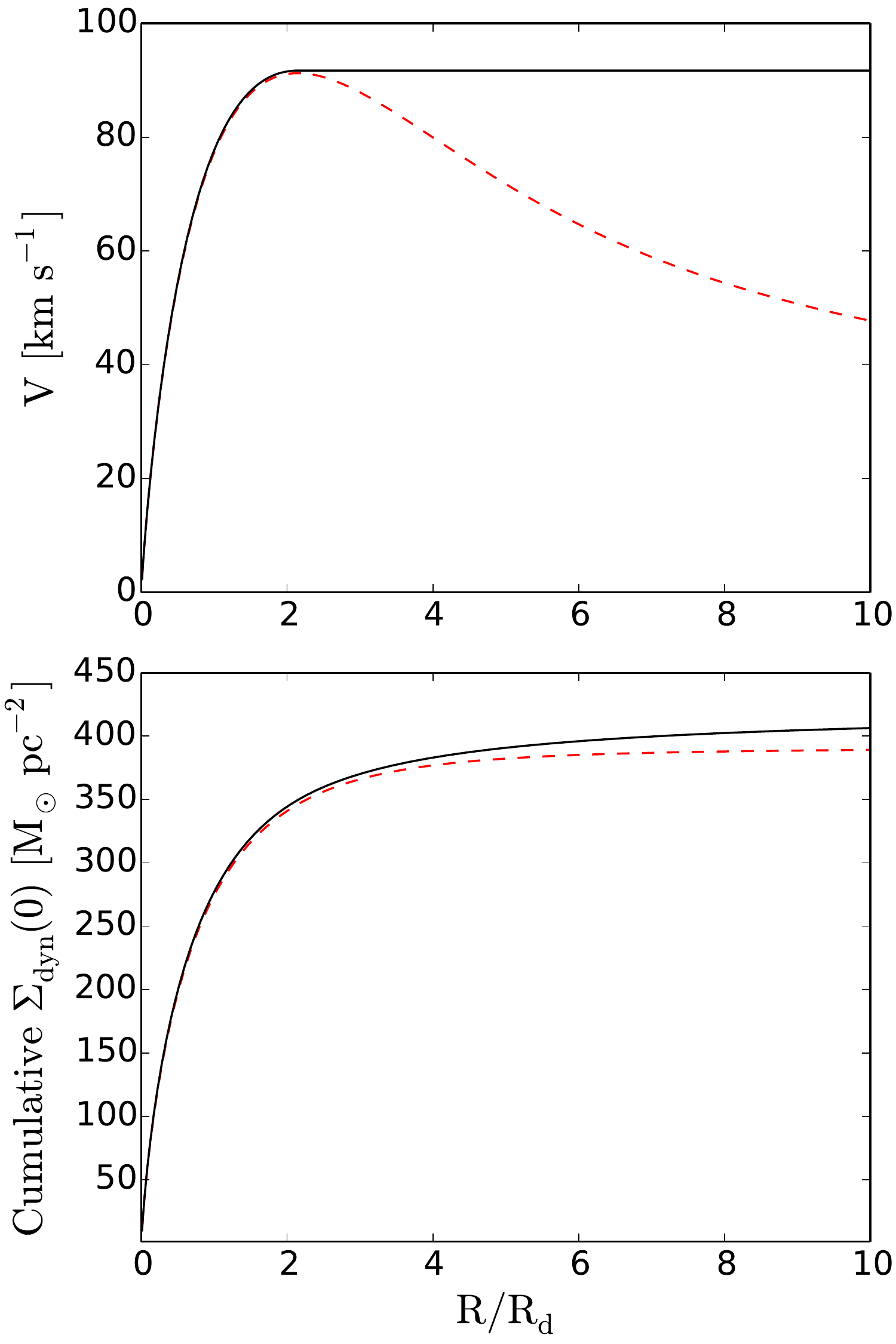}
\caption{\textit{Top panel}: The rotation curve of a self-gravitating exponential disk (dashed line) and a rotation curve that becomes flat beyond 2.2 $R_{\rm d}$ (solid line). \textit{Bottom panel:} The integral value of Eq.\,\ref{Eq:Toomre} between 0 and $R/R_{\rm d}$ for the two models. The resulting \sdyn\ differ by less than 5$\%$ because the addends in Eq.\,\ref{Eq:Toomre} are almost negligible beyond $\sim$3 $R_{\rm d}$.}
\label{Fig:Ex}
\end{figure}
\subsection{Central Dynamical Density}\label{sec:SigmaDyn}

Equation\,\ref{Eq:Toomre} is strictly valid for self-gravitating disks of zero thickness \citep{Toomre1963}. It nevertheless provides a good approximation for real galaxies. In Figure\,\ref{Fig:Ex} we show two simple models: (i) the rotation curve of a razor-thin exponential disk with stellar mass of $10^{10}$ $M_{\odot}$ and scale length $R_{\rm d} = 2$ kpc (for which Eq.\,\ref{Eq:Toomre} holds exactly), and (ii) a realistic rotation curve that becomes flat beyond 2.2 $R_{\rm d}$ due to the DM contribution \citep[e.g.,][]{vanAlbada1986}. The difference in \sdyn\ between these two models is less than 5$\%$ because the addends $V^{2}/R^{2}$ in Eq.\,\ref{Eq:Toomre} are almost negligible beyond $\sim$3$R_{\rm d}$, decreasing as $R^{-2}$ or faster. The value of \sdyn\ is driven by the inner rising portion of the rotation curve. Equation\,\ref{Eq:Toomre} is a good approximation for maximal disks. 

We now consider the extreme case of a galaxy with null disk contribution and a spherical pseudo-isothermal halo
\begin{equation}
 \rho_{\rm ISO}(r) = \dfrac{\rho_{0}}{1 + (r/r_{\rm c})^2},
\end{equation}
where $\rho_{0}$ and $r_{\rm c}$ are the central 3D density and core radii, respectively. This gives
\begin{equation}
 V^{2}_{\rm ISO}(r) = 4\pi G \rho_{0}r_{\rm c}^{2} \bigg[ 1 - \dfrac{r_{\rm c}}{r} \tan^{-1}\bigg(\dfrac{r}{r_{\rm c}}\bigg) \bigg].
\end{equation}
Using Eq.\,\ref{Eq:Toomre} and defining $x = r/r_{\rm c}$, we derive
\begin{equation}\label{Eq:ISO1}
 \Sigma_{\rm dyn}(0) = 2\rho_{0} r_{\rm c}\int_{0}^{\infty} \bigg[ \dfrac{1}{x^2}-\frac{\tan^{-1}(x)}{x^3}\bigg] dx = \dfrac{\pi}{2} \rho_{0} r_{\rm c}.
\end{equation}
This is half the projected 2D central surface density:
\begin{equation}\label{Eq:ISO2}
 \Sigma_{\rm ISO}(0) = \int_{-\infty}^{\infty}\rho_{\rm ISO}(R=0, z) dz = \pi \rho_{0} r_{\rm c}.
\end{equation}
This difference is simply due to the assumed geometry: razor-thin disk or sphere (cf.\,Eq.\,\ref{Eq:Toomre2}). For submaximal disks, therefore, Eq.\,\ref{Eq:Toomre} may underestimate $\Sigma_{\rm dyn}(0)$ by less than a factor of 2. This is a reasonable approximation when comparing surface densities over 4\,dex.

For disks with finite thickness \sdyn\ increases by a factor $(1+q_{0})$, where $q_{0}$ is the intrinsic axial ratio (Agris Kalnajs, private communication). Moreover, since observed rotation curves provide discrete measurements within radial rings, we rewrite Eq.\,\ref{Eq:Toomre} as
\begin{equation}\label{Eq:Toomre2}
  \Sigma_{\rm dyn}(0) = \dfrac{1 + q_{0}}{2\pi G} \sum_{j} \dfrac{V_{j}^{2}}{R_{j}^{2}} \, \Delta R_{j},
\end{equation}
where $\Delta R_{j}$ is the width of the $j$-th radial ring. We use model rotation curves (Figure\,\ref{Fig:Ex}) to test the effect of radial sampling and find that it is modest. If the rising portion of the rotation curve is sampled with only 4 points, \sdyn\ is underestimated by only $\sim$35$\%$. This is an extreme example: the majority of SPARC galaxies have better spatial sampling.

The intrinsic axial ratio of galaxies may vary with stellar mass $M_{\star}$ \citep{Sanchez2010} or Hubble type $T$ \citep{Yuan2004}. We adopt the following parabolic relation:
\begin{equation}
 q_{0} = 0.0625 \log(M_{\star})^{2} - 1.125\log(M_{\star}) + 5.2125,
\end{equation}
which reaches a minimum value of $q_{0}=0.15$ at $M_{\star} = 10^{9} M_{\odot}$ and gives $q_{0} = 0.4$ at $M_{\star} = 10^{7} M_{\odot}$ and $M_{\star} = 10^{11} M_{\odot}$ (see Figure 1 of \citealt{Sanchez2010}). We also tried the $q_{0}-T$ relations of \citet{Yuan2004} and a constant $q_{0} = 0.2$: we find only minor differences although the scatter on the \sdyn$-$\sstar\ relation slightly increases. Note that $q_{0}$ may vary with radius but here we are only interested in the typical value for the inner galaxy regions.

The error on $\Sigma_{\rm dyn}$ is estimated as
\begin{equation}
\begin{split}
 \delta_{\Sigma_{\rm dyn}(0)}^{2} &= \dfrac{1 + q_{0}}{2\pi G}\sum_{j} \bigg[ \dfrac{2 V_{j}^2 \Delta R_{j} }{R_{j}^2} \dfrac{\delta_{V_j}}{V_j}\bigg]^{2}+ \bigg[\Sigma_{\rm dyn}(0) \dfrac{\delta_{D}}{D}\bigg]^{2} \\
                               &+ \bigg[2\Sigma_{\rm dyn}(0)\dfrac{\delta_{i}}{\tan(i)}\bigg]^{2} + \bigg[\Sigma_{\rm dyn}(0) \dfrac{\delta_{q_{0}}}{1+q_{0}}\bigg]^{2},
\end{split}
\end{equation}
where we consider the errors on each individual velocity point ($\delta_{V_{j}}$), disk inclination ($\delta_{i}$), galaxy distance ($\delta_{D}$), and axial ratio ($\delta_{q_{0}}$). We refer to Paper I for the errors on the rotation velocities, galaxy distance, and disk inclination. For $q_{0}$ we assume a conservative error of 50$\%$. The \textit{formal} average error on \sdyn\ is 0.14 dex.

\begin{figure*}[thb]
\centering
\includegraphics[width=\textwidth]{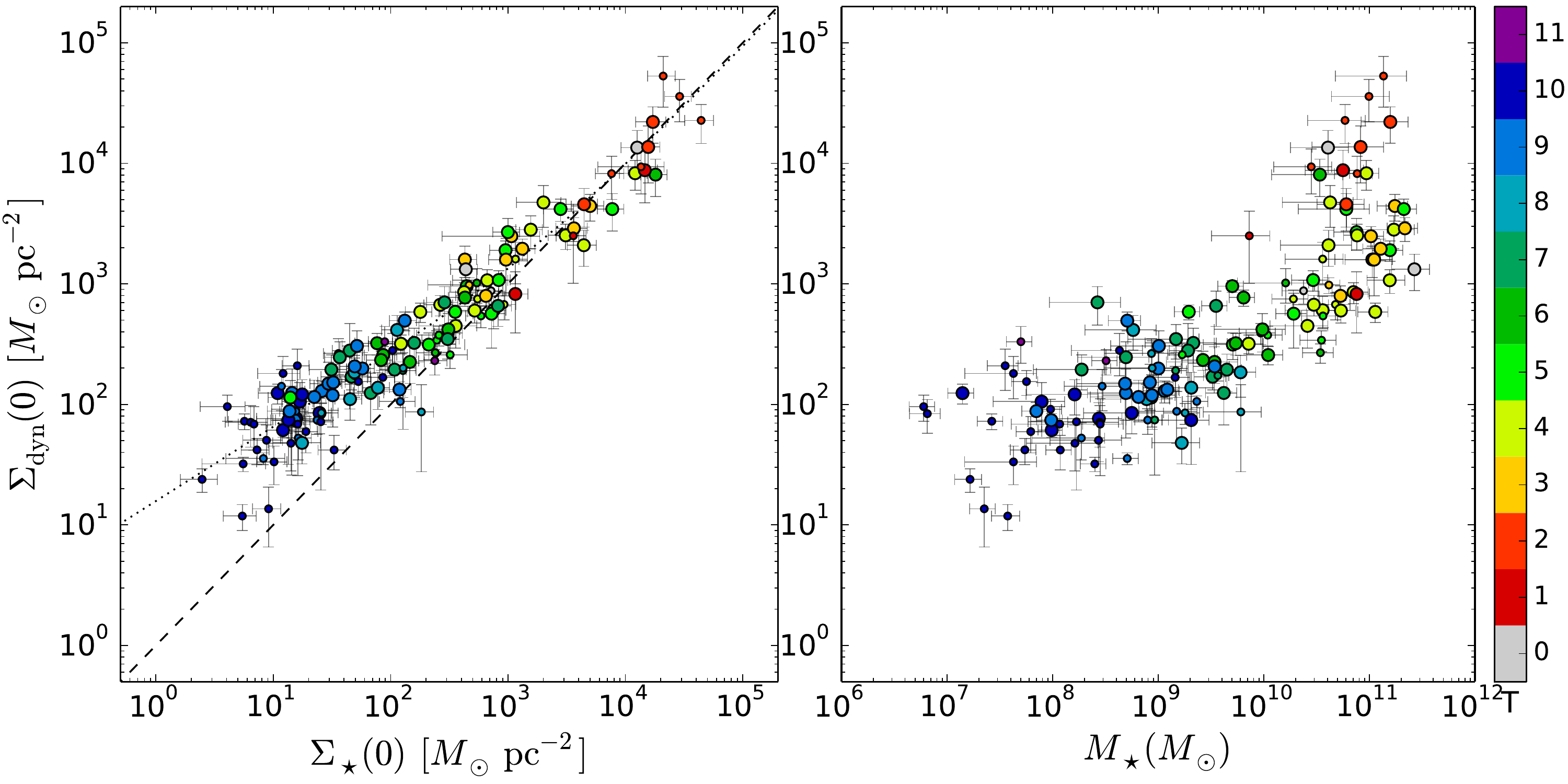}
\caption{Central dynamical surface density versus central stellar surface density (\textit{left}) and total stellar mass (\textit{right}). Galaxies are color-coded by numerical Hubble type. Large and small symbols indicate galaxies with quality flag $Q=1$ and $Q=2$, respectively. In the left panel, the dashed and dotted lines indicate the 1:1 relation and a double power-law fit, respectively.}
\label{fig:Central}
\end{figure*}
\subsection{Central Stellar Density}\label{sec:SigmaStar}

We estimate the central stellar density using [3.6] surface brightness profiles from Paper I. These profiles often have finer radial samplings than \hiA/H$\alpha$ rotation curves due to the higher spatial resolution of $Spitzer$ [3.6] images. To properly compare \sstar\ with \sdyn\, we calculate the average surface brightness within the first measured point of the rotation curve (ranging from $\sim$1$''$ to $\sim$15$''$ depending on the galaxy). Hence, we effectively smooth the [3.6] profiles to the \hiA/H$\alpha$ spatial resolution.

We convert surface brightnesses to stellar surface densities using a solar absolute magnitude of 3.24 \citep{Oh2008} and a constant stellar mass-to-light ratio at 3.6 $\mu$m ($\Upsilon_{[3.6]}$). \citet{Lelli2016b} find that the baryonic Tully-Fisher relation is very tight for a fixed \ml, hence the actual \ml\ cannot vary wildly among galaxies \citep[see also][]{McGaugh2015}. Here we distinguish between bulge and disk components using the non-parametric decompositions from Paper\,I for 32 early-type spirals (S0 to Sb). \citet{Schombert2014a} build stellar population synthesis (SPS) models with a \citet{Chabrier2003} initial mass function (IMF), finding $\Upsilon_{[3.6]} \simeq 0.5$ $M_{\odot}/L_{\odot}$ for star-forming disks and $\Upsilon_{[3.6]} \simeq 0.7$ $M_{\odot}/L_{\odot}$ for bulges. These values are used here. They are consistent with different SPS models \citep{Meidt2014, McGaugh2014}, agree with resolved stellar populations in the LMC \citep{Eskew2012}, and provides sensible gas fractions (Paper\,I). For the disk component, the central surface brightness is corrected to face-on view assuming that internal extinction is negligible in the near IR \citep[e.g.,][]{Verheijen2001b}.

The error on \sstar\ is estimated as
\begin{equation}
 \delta_{\Sigma_{\star}(0)} = \sqrt{[0.25 \Sigma_{\star}(0)]^{2} + [\Sigma_{\star}(0) \tan(i) \delta_{i}]^{2}}.
\end{equation}
The factor 0.25 represents a scatter of $\sim$0.11 dex on \ml\ as suggested by SPS models \citep{McGaugh2014, Meidt2014, Schombert2014a}. The \textit{formal} average error on \sstar\ is 0.14 dex.

\begin{figure*}[thb]
\centering
\includegraphics[width=\textwidth]{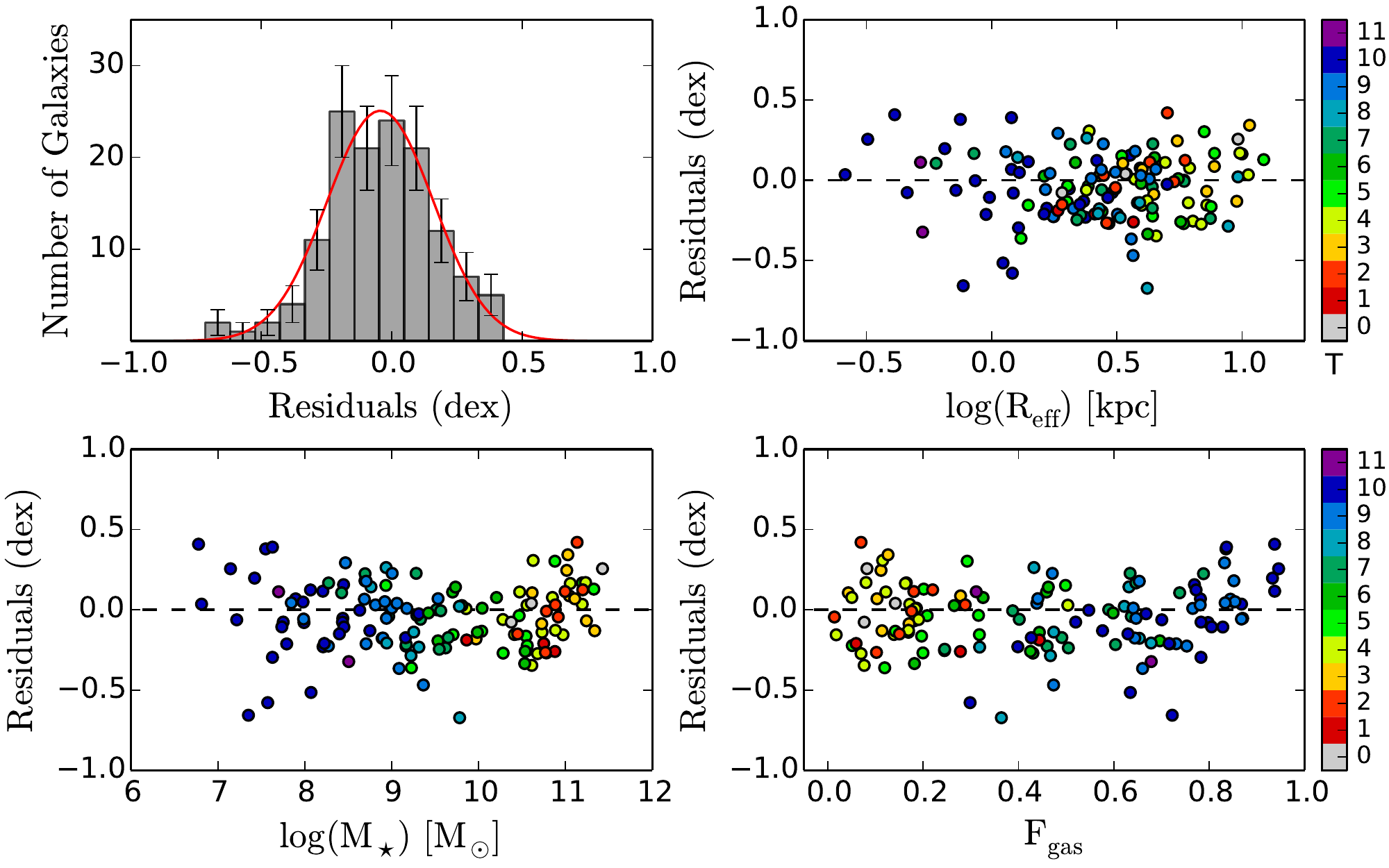}
\caption{Residuals around the central density relation obtained by subtracting Eq.\,\ref{Eq:DoublePower} with the parameters in Eq.\,\ref{Eq:Para}. The top-left panel shows a histogram with Poissonian ($\sqrt{N}$) errorbars. The red line shows a Gaussian fit. The other panels show the residuals versus effective radius (\textit{top-right}), stellar mass (\textit{bottom-left}), and gas fraction (\textit{bottom-right}). Galaxies are color-coded by numerical Hubble type.}
\label{fig:CDRres}
\end{figure*}

\section{The Central Density Relation}

\subsection{General Results}\label{sec:Results}

Figure \ref{fig:Central} (left panel) shows \sdyn\ versus \sstar. We find a non-linear relation with a break around $\Sigma_{\star}(0)\simeq 1000$ $M_{\odot}$~pc$^{-2}$. HSB galaxies lie on the 1:1 line, indicating that they are baryon dominated in the inner parts. A few points have unphysical values of $\Sigma_{\star}(0)/\Sigma_{\rm dyn}(0)>1$ but are consistent with 1 within the errors. This is expected to happen since the actual \ml\ may scatter to values higher/lower than the adopted mean \ml.

LSB galaxies systematically deviate from the 1:1 line, indicating that they are increasingly DM dominated in the inner regions. This is another manifestation of the long-standing fine-tuning problem between baryonic fraction and surface brightness in disk galaxies \citep{Zwaan1995, McGaugh1998}. Recovering a single linear relation requires that LSB galaxies have systematically higher $\Upsilon_{\star}$ than HSB ones. This is unrealistic since LSB disks have young stellar populations and low metallicities \citep{Impey1997, Schombert2014a, Schombert2014b, Schombert2015}. Even though Toomre's formula may break down for non-self-gravitating LSB galaxies, the relation remains remarkably tight.

We point out that using $\Sigma_{\rm bar}(0) = \Sigma_{\rm gas}(0)+\Sigma_{\star}(0)$ would not drastically change the overall relation. The central \hi surface density of disk galaxies is generally between 3 and 12 $M_{\odot}$~pc$^{-2}$ \citep{deBlok1996a, Swaters2002, Lelli2014a}, hence atomic gas gives a negligible contribution to $\Sigma_{\rm bar}(0)$ with the possible exception of extreme LSB galaxies. The effect of molecules is more difficult to quantify because CO gas is often undetected in low-mass LSB galaxies and the CO-to-H$_2$ conversion factor may vary with metallicity or other ISM properties \citep{Schruba2012}. In general, we expect LSB galaxies to have low H$_2$ surface densities ($\lesssim$10 M$_{\odot}$ pc$^{-2}$) given their low star-formation rate densities \citep{Mihos1999}. On the other hand, HSB galaxies can show high H$_2$ surface densities but $\Sigma_{\rm bar}$ is nevertheless dominated by stars near the centre \citep{Frank2016}.

Figure \ref{fig:Central} also shows \sdyn\ versus $M_{\star}$ (right panel). Clearly, \sdyn\ is more closely related to stellar density than stellar mass. At low masses ($M_{\star} \lesssim 10^{10} M_{\odot}$), there is a broad trend that derives from the well-known $\Sigma_{\star} - M_{\star}$ relation (e.g., Paper I). At high masses, however, this broad trend breaks down: galaxies with the same $M_{\star}$ can show differences in \sdyn\ up to 2 dex. This seems to be driven by bulges: at a fixed $M_{\star}$ galaxies with high \sdyn\ tend to be Sa/Sab ($T=1-2$).

\subsection{Fits and Residuals}\label{sec:Fits}

We fit a generic double power-law:
\begin{equation}\label{Eq:DoublePower}
 \Sigma_{\rm dyn}(0) = \Sigma_{0} \bigg[ 1 + \dfrac{\Sigma_{\star}(0)}{\Sigma_{\rm crit}}\bigg]^{\alpha-\beta} \bigg[\dfrac{\Sigma_{\star}(0)}{\Sigma_{\rm crit}}\bigg]^{\beta},
\end{equation}
where $\alpha$ and $\beta$ are asymptotic slopes for $\Sigma_{\star}(0) \gg \Sigma_{\rm crit}$ and $\Sigma_{\star}(0) \ll \Sigma_{\rm crit}$, respectively. We fit the data using the \emph{Python} orthogonal distance regression algorithm (scipy.odr), considering errors in both variables. The fit is poorly constrained when all four parameters are allowed to vary. This is likely due to the paucity of very HSB galaxies in our sample. Fixing $\alpha = 1$, we find
\begin{equation}\label{Eq:Para}
\begin{split}
&\beta = 0.62 \pm 0.04\\
&\Sigma_{0} = 1468 \pm 866 \quad [M_{\odot} \, \mathrm{pc}^{-2}]\\
&\Sigma_{\rm crit} = 1571 \pm 1137 \quad [M_{\odot} \, \mathrm{pc}^{-2}]
\end{split}
 \end{equation}
Note that $\Sigma_{0} \simeq \Sigma_{\rm crit}$: this implies that the ``intercept'' is consistent with the 1:1 line at high-densities. Leaving $\alpha$ as a free parameter and imposing $\Sigma_{0} = \Sigma_{\rm crit}$, we find 
\begin{equation}
\begin{split}
&\alpha = 0.97 \pm 0.06\\
&\beta = 0.61 \pm 0.04\\
&\Sigma_{0} = \Sigma_{\rm crit} = 1271 \pm 463 \quad [M_{\odot} \, \mathrm{pc}^{-2}]
\end{split}
\end{equation}
Clearly, the high-end of the relation is fully consistent with the 1:1 line in both slope and intercept.

The observed scatter around these fits is 0.21 dex. This is largely driven by observational uncertainties. On average, we have 
\begin{equation}
\sqrt{\delta_{\Sigma_{\rm dyn}(0)}^2 + \delta_{\Sigma_{\star}(0)}^2} = \sqrt{0.14^2 + 0.14^2} = 0.20 \, \rm{dex}. 
\end{equation}
This formal error neglects that different galaxies are observed at different linear resolutions (in kpc), which may likely account for the small remaining scatter.

The residuals are represented by a histogram in Figure \ref{fig:CDRres} (top-left). A Gaussian function provides a good fit but it is slightly offset from zero (by $-$0.04 dex): this is consistent with Poisson noise given the relatively low number of galaxies in each bin. The standard deviation from the Gaussian fit is 0.20 dex in agreement with the measured scatter of 0.21 dex. In the other panels of Figure \ref{fig:CDRres}, we plot the residuals against several structural parameters: the stellar effective radius $R_{\rm eff}$ (top-right), the total stellar mass $M_{\star}$ (bottom-left), and the gas fraction $F_{\rm gas} = M_{\rm gas}/M_{\rm bar}$ (bottom-right). The residuals display no correlation with any of these quantities: the Pearson's, Spearman's, and Kendall's coefficients are consistently between 0 and 0.1. These structural parameters do not play a role in setting \sdyn. 

\section{Discussion}

In this Letter we employ a formula from \citet{Toomre1963} to estimate the central dynamical density \sdyn\ in disk galaxies. We consider 135 galaxies (S0 to dIrr) from the SPARC dataset, which have high-quality \hiA/H$\alpha$ rotation curves and [3.6] surface brightness profiles (Paper\,I). We find that \sdyn\ correlates with \sstar\ over 4 dex. This central density relation can be described by a double power-law. The observed scatter is small ($\sim$0.2 dex) and largely driven by observational uncertainties. The residuals show no correlations with other galaxy properties. Since $\Sigma_{\rm dyn}(0)$ is mostly driven by the inner rising portion of the rotation curve (Sect.\,\ref{sec:SigmaDyn}), this scaling relation is closely connected to the one from \citet{Lelli2013}: the new relation has the advantage of linking baryonic and dynamical surface densities in a more direct and quantitative way.

\subsection{The case for maximum disks in HSB galaxies}\label{sec:HSB}

In this work, we estimate the central stellar densities using $\Upsilon_{[3.6]} = 0.5$ $M_{\odot}/L_{\odot}$ for disks and $\Upsilon_{[3.6]} = 0.7$ $M_{\odot}/L_{\odot}$ for bulges (in 32 early-type spirals). These values are derived from SPS models with a \citet{Chabrier2003} IMF \citep{Schombert2014a}. Using two different \ml\ encapsulates the different evolutionary histories of star-forming disks (dominated by young stellar populations) and bulges (dominated by old stellar populations). Apart from this minor detail, the \textit{shape} of the central density relation does not depend on the specific choice of \ml. 

Alternative normalizations of \ml\ (e.g., due to different IMFs or SPS models) would horizontally shift the relation and change the values of $\Sigma_{0}$ and $\Sigma_{\rm crit}$ in Equation~\ref{Eq:DoublePower}, but leave the slopes $\alpha$ and $\beta$ unaffected. In particular, the linearity of the relation at high densities does not depend on \ml\ and strongly supports the concept of baryonic dominance in the inner parts of HSB galaxies. \textit{If HSB galaxies were strongly sub-maximal there would be no reason to have a relation with a high-end slope of 1.} Since Eq.~\ref{Eq:Toomre} is valid for self-gravitating disks, the adopted normalization of \ml\ seems natural as it places HSB galaxies on the 1:1 line.

\subsection{The baryon-DM coupling in LSB galaxies}\label{sec:LSB}

The high-end of the central density relation can be trivially explained if baryons dominate the inner parts of HSB galaxies. The low-end, however, is puzzling. For DM-dominated LSB galaxies, Toomre's equation may underestimate \sdyn\ up to a factor of 2 depending on the actual DM distribution and disk contribution (Sect.\,\ref{sec:SigmaDyn}). It nevertheless returns a tight correlation. Since Eq.\,\ref{Eq:Toomre} is obtained by solving Poisson's equation for a flattened mass distribution, one may speculate that the baryonic and total gravitational potentials describe similar isopotential surfaces, in line with the predictions of Modified Newtonian Dynamics \citep{Milgrom1983}. In any case, baryons and DM must be tightly coupled \citep[see also][]{Sancisi2004, Swaters2012, Lelli2013}.

In a $\Lambda$CDM context, the slowly-rising rotation curves of LSB galaxies are in contradiction with the cuspy DM haloes predicted by cosmological $N$-body simulations \citep[e.g.,][]{deBlok2001, Gentile2004}. Stellar feedback is frequently invoked to redistribute the primordial DM distribution and transform the predicted cusps into the observed cores \citep[e.g.,][]{Governato2010, Teyssier2013, DiCintio2014, Madau2014}. Baryon-modified DM haloes provide good fits to the observed rotation curves and simultaneously recover $\Lambda$CDM scaling relations \citep{Katz2016}. This stochastic process, however, needs to reproduce the central density relation with virtually \textit{no} intrinsic scatter. Moreover, one may expect that the residuals correlate with galaxy mass, size, or gas fraction, but no such correlation is observed (Sect.\,\ref{sec:Fits}). These are open challenges for $\Lambda$CDM models of galaxy formation.

\subsection{Comparison with the DiskMass survey}

A similar central density relation has been reported by \citet{Swaters2014} using 30 face-on galaxies from the DiskMass survey \citep[DMS,][]{Bershady2010}. For these objects, $\Sigma_{\rm dyn}(0)$ is estimated measuring the stellar velocity dispersion and assuming that the disk is self-gravitating. This approach also requires assumptions on the shape of the velocity dispersion tensor and on the vertical mass distribution (crucially the disk scale height). \citet{Swaters2014} finds that $\Sigma_{\rm dyn}(0)$ correlates with $\mu_{0}$ over less than 1 dex (from $\sim$400 to $\sim$2000 $L_{\odot}$ pc$^{-2}$): the five LSB galaxies in the DMS sample fall above their fitted relation. Our relation is much more extended thanks to the excellent dynamic range of SPARC. Our sample is $\sim$5 times larger than that of \citet{Swaters2014} and more extended by $\sim$1 and $\sim$2 dex at the high and low density ends, respectively. In particular, we show that the ``deviating'' LSB galaxies in \citet{Swaters2014} were hinting at a genuine change in slope rather than a break-down of the central density relation.

Interestingly, \citet{Swaters2014} find a \textit{linear} relation for HSB galaxies with small intrinsic scatter. This is consistent with our results in Sect.\,\ref{sec:Fits}. The normalization, however, differs by a factor of $\sim$2. As we discussed in Sect.\,\ref{sec:HSB}, a central density relation with a slope of 1 is a natural outcome for maximum disks and confirms $-$a posteriori$-$ the hypothesis of self-gravitating disks, which is implicit in both our Eq.~\ref{Eq:Toomre} and Eq.~1 of \citet{Swaters2014}. A constant shift from the 1:1 line may hint at systematics. For example, \citet{Aniyan2016} point out that the scale-heights employed by the DMS are representative of old stellar populations, whereas the velocity dispersions come from integrated light measurements with contributions from K-giants of different ages. In the solar vicinity, \citet{Aniyan2016} find that young K-giants have significantly smaller velocity dispersion and scale-height than old K-giants, hence the DMS may have systematically underestimated the dynamical surface densities by a factor of $\sim$2. If true, the relation from \citet{Swaters2014} would shift in the vertical direction and be fully consistent with our relation.

\section{Conclusions}

In this Letter we establish a scaling relation between the central dynamical density and the central stellar density of disk galaxies. HSB galaxies are consistent with unity, suggesting that they are self-gravitating and baryon dominated in the inner parts. LSB galaxies systematically deviate from the 1:1 line, indicating that the DM contribution progressively increases but remains tightly coupled to the baryonic one. This central density relation represents a key testbed for cosmological models of galaxy formation.

\acknowledgments

We are indebted to Agris Kalnajs for making us aware of the ``widely forgotten'' formula from Toomre (1963). This publication was made possible through the support of a grant from the John Templeton  Foundation. The opinions expressed in this publication are those of the authors and do not necessarily reflect the views of the John Templeton Foundation.


\begin{thebibliography}{}
\expandafter\ifx\csname natexlab\endcsname\relax\def\natexlab#1{#1}\fi

\bibitem[{{Aniyan} {et~al.}(2016){Aniyan}, {Freeman}, {Gerhard}, {Arnaboldi},
  \& {Flynn}}]{Aniyan2016}
{Aniyan}, S., {Freeman}, K.~C., {Gerhard}, O.~E., {Arnaboldi}, M., \& {Flynn},
  C. 2016, \mnras, 456, 1484

\bibitem[{{Bershady} {et~al.}(2010){Bershady}, {Verheijen}, {Swaters},
  {Andersen}, {Westfall}, \& {Martinsson}}]{Bershady2010}
{Bershady}, M.~A., {Verheijen}, M.~A.~W., {Swaters}, R.~A., {et~al.} 2010,
  \apj, 716, 198

\bibitem[{{Chabrier}(2003)}]{Chabrier2003}
{Chabrier}, G. 2003, \apjl, 586, L133

\bibitem[{{de Blok} \& {McGaugh}(1996)}]{deBlok1996b}
{de Blok}, W.~J.~G., \& {McGaugh}, S.~S. 1996, \apjl, 469, L89

\bibitem[{{de Blok} {et~al.}(2001){de Blok}, {McGaugh}, \&
  {Rubin}}]{deBlok2001}
{de Blok}, W.~J.~G., {McGaugh}, S.~S., \& {Rubin}, V.~C. 2001, AJ, 122, 2396

\bibitem[{{de Blok} {et~al.}(1996){de Blok}, {McGaugh}, \& {van der
  Hulst}}]{deBlok1996a}
{de Blok}, W.~J.~G., {McGaugh}, S.~S., \& {van der Hulst}, J.~M. 1996, \mnras,
  283, 18

\bibitem[{{Di Cintio} {et~al.}(2014){Di Cintio}, {Brook}, {Macci{\`o}},
  {Stinson}, {Knebe}, {Dutton}, \& {Wadsley}}]{DiCintio2014}
{Di Cintio}, A., {Brook}, C.~B., {Macci{\`o}}, A.~V., {et~al.} 2014, \mnras,
  437, 415

\bibitem[{{Erroz-Ferrer} {et~al.}(2016){Erroz-Ferrer}, {Knapen}, {Leaman},
  {D{\'{\i}}az-Garc{\'{\i}}a}, {Salo}, {Laurikainen}, {Querejeta}, {Mu
  noz-Mateos}, {Athanassoula}, {Bosma}, {Comer{\'o}n}, {Elmegreen}, \&
  {Mart{\'{\i}}nez-Valpuesta}}]{ErrozFerrer2016}
{Erroz-Ferrer}, S., {Knapen}, J.~H., {Leaman}, R., {et~al.} 2016, \mnras,
  arXiv:1602.02789

\bibitem[{{Eskew} {et~al.}(2012){Eskew}, {Zaritsky}, \& {Meidt}}]{Eskew2012}
{Eskew}, M., {Zaritsky}, D., \& {Meidt}, S. 2012, \aj, 143, 139

\bibitem[{{Frank} {et~al.}(2015){Frank}, {de Blok}, {Walter}, {Leroy}, \&
  {Carignan}}]{Frank2016}
{Frank}, B.~S., {de Blok}, W.~J.~G., {Walter}, F., {Leroy}, A., \& {Carignan},
  C. 2015, ArXiv e-prints, arXiv:1512.01367

\bibitem[{{Freeman}(1970)}]{Freeman1970}
{Freeman}, K.~C. 1970, ApJ, 160, 811

\bibitem[{{Gentile} {et~al.}(2004){Gentile}, {Salucci}, {Klein}, {Vergani}, \&
  {Kalberla}}]{Gentile2004}
{Gentile}, G., {Salucci}, P., {Klein}, U., {Vergani}, D., \& {Kalberla}, P.
  2004, \mnras, 351, 903

\bibitem[{{Governato} {et~al.}(2010){Governato}, {Brook}, {Mayer}, {Brooks},
  {Rhee}, {Wadsley}, {Jonsson}, {Willman}, {Stinson}, {Quinn}, \&
  {Madau}}]{Governato2010}
{Governato}, F., {Brook}, C., {Mayer}, L., {et~al.} 2010, \nat, 463, 203

\bibitem[{{Impey} \& {Bothun}(1997)}]{Impey1997}
{Impey}, C., \& {Bothun}, G. 1997, \araa, 35, 267

\bibitem[{{Katz} {et~al.}(2016){Katz}, {Lelli}, {McGaugh}, {Di Cintio},
  {Brooks}, \& {Schombert}}]{Katz2016}
{Katz}, H., {Lelli}, F., {McGaugh}, S.S., {et~al.} 2016, arXiv:1605.05971

\bibitem[{{Lelli}(2014)}]{Lelli2014}
{Lelli}, F. 2014, Galaxies, 2, 292

\bibitem[{{Lelli} {et~al.}(2013){Lelli}, {Fraternali}, \&
  {Verheijen}}]{Lelli2013}
{Lelli}, F., {Fraternali}, F., \& {Verheijen}, M. 2013, \mnras, 433, L30

\bibitem[{{Lelli} {et~al.}(2016{\natexlab{a}}){Lelli}, {McGaugh}, \&
  {Schombert}}]{Lelli2016a}
{Lelli}, F., {McGaugh}, S.~S., \& {Schombert}, J.~M. 2016{\natexlab{a}}, ArXiv
  e-prints, arXiv:1606.09251

\bibitem[{{Lelli} {et~al.}(2016{\natexlab{b}}){Lelli}, {McGaugh}, \& {Schombert}}]{Lelli2016b}
{Lelli}, F., {McGaugh}, S.~S., \& {Schombert}, J.~M. 2016{\natexlab{b}}, \apjl, 816, L14

\bibitem[{{Lelli} {et~al.}(2014){Lelli}, {Verheijen}, \&
  {Fraternali}}]{Lelli2014a}
{Lelli}, F., {Verheijen}, M., \& {Fraternali}, F. 2014, \aap, 566, A71

\bibitem[{{Madau} {et~al.}(2014){Madau}, {Shen}, \& {Governato}}]{Madau2014}
{Madau}, P., {Shen}, S., \& {Governato}, F. 2014, \apjl, 789, L17

\bibitem[{{Martinsson} {et~al.}(2013){Martinsson}, {Verheijen}, {Westfall},
  {Bershady}, {Andersen}, \& {Swaters}}]{Martinsson2013}
{Martinsson}, T.~P.~K., {Verheijen}, M.~A.~W., {Westfall}, K.~B., {et~al.}
  2013, \aap, 557, A131

\bibitem[{{McGaugh}(2012)}]{McGaugh2012}
{McGaugh}, S.~S. 2012, \, 143, 40

\bibitem[{{McGaugh} \& {de Blok}(1998)}]{McGaugh1998}
{McGaugh}, S.~S., \& {de Blok}, W.~J.~G. 1998, \apj, 499, 41

\bibitem[{{McGaugh} \& {Schombert}(2014)}]{McGaugh2014}
{McGaugh}, S.~S., \& {Schombert}, J.~M. 2014, \aj, 148, 77

\bibitem[{{McGaugh} \& {Schombert}(2015)}]{McGaugh2015}
---. 2015, \apj, 802, 18

\bibitem[{{Meidt} {et~al.}(2014){Meidt}, {Schinnerer}, {van de Ven},
  {Zaritsky}, {Peletier}, {Knapen}, {Sheth}, {Regan}, {Querejeta},
  {Mu{\~n}oz-Mateos}, {Kim}, {Hinz}, {Gil de Paz}, {Athanassoula}, {Bosma},
  {Buta}, {Cisternas}, {Ho}, {Holwerda}, {Skibba}, {Laurikainen}, {Salo},
  {Gadotti}, {Laine}, {Erroz-Ferrer}, {Comer{\'o}n}, {Men{\'e}ndez-Delmestre},
  {Seibert}, \& {Mizusawa}}]{Meidt2014}
{Meidt}, S.~E., {Schinnerer}, E., {van de Ven}, G., {et~al.} 2014, \apj, 788,
  144

\bibitem[{{Mihos} {et~al.}(1999){Mihos}, {Spaans}, \& {McGaugh}}]{Mihos1999}
{Mihos}, J.~C., {Spaans}, M., \& {McGaugh}, S.~S. 1999, \apj, 515, 89

\bibitem[{{Milgrom}(1983)}]{Milgrom1983}
{Milgrom}, M. 1983, ApJ, 270, 371

\bibitem[{{Oh} {et~al.}(2008){Oh}, {de Blok}, {Walter}, {Brinks}, \&
  {Kennicutt}}]{Oh2008}
{Oh}, S.-H., {de Blok}, W.~J.~G., {Walter}, F., {Brinks}, E., \& {Kennicutt},
  Jr., R.~C. 2008, \aj, 136, 2761

\bibitem[{{S{\'a}nchez-Janssen} {et~al.}(2010){S{\'a}nchez-Janssen},
  {M{\'e}ndez-Abreu}, \& {Aguerri}}]{Sanchez2010}
{S{\'a}nchez-Janssen}, R., {M{\'e}ndez-Abreu}, J., \& {Aguerri}, J.~A.~L. 2010,
  \mnras, 406, L65

\bibitem[{{Sancisi}(2004)}]{Sancisi2004}
{Sancisi}, R. 2004, in IAU Symposium, Vol. 220, Dark Matter in Galaxies, ed.
  S.~{Ryder}, D.~{Pisano}, M.~{Walker}, \& K.~{Freeman}, 233

\bibitem[{{Schombert} \& {McGaugh}(2014{\natexlab{a}})}]{Schombert2014a}
{Schombert}, J., \& {McGaugh}, S. 2014{\natexlab{a}}, \pasa, 31, 36

\bibitem[{{Schombert} \& {McGaugh}(2014{\natexlab{b}})}]{Schombert2014b}
{Schombert}, J.~M., \& {McGaugh}, S. 2014{\natexlab{b}}, \pasa, 31, 11

\bibitem[{{Schombert} \& {McGaugh}(2015)}]{Schombert2015}
{Schombert}, J.~M., \& {McGaugh}, S.~S. 2015, \aj, 150, 72

\bibitem[{{Schruba} {et~al.}(2012){Schruba}, {Leroy}, {Walter}, {Bigiel},
  {Brinks}, {de Blok}, {Kramer}, {Rosolowsky}, {Sandstrom}, {Schuster},
  {Usero}, {Weiss}, \& {Wiesemeyer}}]{Schruba2012}
{Schruba}, A., {Leroy}, A.~K., {Walter}, F., {et~al.} 2012, \aj, 143, 138

\bibitem[{{Swaters} {et~al.}(2014){Swaters}, {Bershady}, {Martinsson},
  {Westfall}, {Andersen}, \& {Verheijen}}]{Swaters2014}
{Swaters}, R.~A., {Bershady}, M.~A., {Martinsson}, T.~P.~K., {et~al.} 2014,
  \apjl, 797, L28

\bibitem[{{Swaters} {et~al.}(2012){Swaters}, {Sancisi}, {van der Hulst}, \&
  {van Albada}}]{Swaters2012}
{Swaters}, R.~A., {Sancisi}, R., {van der Hulst}, J.~M., \& {van Albada}, T.~S.
  2012, MNRAS, 425, 2299

\bibitem[{{Swaters} {et~al.}(2002){Swaters}, {van Albada}, {van der Hulst}, \&
  {Sancisi}}]{Swaters2002}
{Swaters}, R.~A., {van Albada}, T.~S., {van der Hulst}, J.~M., \& {Sancisi}, R.
  2002, \aap, 390, 829

\bibitem[{{Teyssier} {et~al.}(2013){Teyssier}, {Pontzen}, {Dubois}, \&
  {Read}}]{Teyssier2013}
{Teyssier}, R., {Pontzen}, A., {Dubois}, Y., \& {Read}, J.~I. 2013, \mnras,
  429, 3068

\bibitem[{{Toomre}(1963)}]{Toomre1963}
{Toomre}, A. 1963, \apj, 138, 385

\bibitem[{{van Albada} \& {Sancisi}(1986)}]{vanAlbada1986}
{van Albada}, T.~S., \& {Sancisi}, R. 1986, Philosophical Transactions of the
  Royal Society of London Series A, 320, 447

\bibitem[{{Verheijen}(1997)}]{Verheijen1997}
{Verheijen}, M.~A.~W. 1997, PhD thesis, University of Groningen

\bibitem[{{Verheijen}(2001)}]{Verheijen2001b}
---. 2001, ApJ, 563, 694

\bibitem[{{Verheijen} \& {Sancisi}(2001)}]{Verheijen2001a}
{Verheijen}, M.~A.~W., \& {Sancisi}, R. 2001, \aap, 370, 765

\bibitem[{{Yuan} \& {Zhu}(2004)}]{Yuan2004}
{Yuan}, Q.-r., \& {Zhu}, C.-x. 2004, \caa, 28, 127

\bibitem[{{Zwaan} {et~al.}(1995){Zwaan}, {van der Hulst}, {de Blok}, \&
  {McGaugh}}]{Zwaan1995}
{Zwaan}, M.~A., {van der Hulst}, J.~M., {de Blok}, W.~J.~G., \& {McGaugh},
  S.~S. 1995, \mnras, 273, L35

\end{thebibliography}

\end{document}